\documentclass[]{aastex631}

\usepackage[utf8]{inputenc}
\usepackage{graphicx}
\usepackage{epstopdf}
\usepackage{verbatim}
\usepackage{amsmath}
\usepackage{multirow}
\usepackage{array}
\newcolumntype{P}[1]{>{\centering\arraybackslash}p{#1}}
\newcolumntype{M}[1]{>{\centering\arraybackslash}m{#1}}

\usepackage[shortlabels]{enumitem}

\usepackage{makecell}

\begin{document}
\title{Close TNO Passages as a Driver of the Origin and Evolution of Ultra-Wide Kuiper Belt Binaries}
\author{Hunter M. Campbell}
\author{Lukas R. Stone}
\author{Nathan A. Kaib}

\affiliation{1HL Dodge Department of Physics \& Astronomy, University of Oklahoma, Norman, OK 73019, USA
}

\begin{abstract}
Within the dynamically cold low inclination portion of the Classical Kuiper Belt, there exists a population of weakly bound binary systems with a number of unusual properties; most notable of which is their extremely wide orbital separations; beyond 7\% of their Hill radii.  The stability and evolution of these Ultra-Wide Trans-Neptunian Binaries (TNBs) have, in the past, been studied extensively under the assumption that the primary evolving mechanisms are interactions between the binary components and impacting Trans-Neptunian Objects (TNOs).  Here, we instead study their evolution as driven by the gravitational perturbations of close passing but non-impacting TNOs.  By simulating these passages, we show that the aggregate effects of encounters over billions of years have a significant effect on Kuiper Belt binary evolution.  Such processes can lead to tight binaries widening significantly over time, approaching and sometimes surpassing the separation of the widest known TNBs.  We also find that the eccentricity and inclination distributions of observed Ultra-Wide TNBs can be sampled from such widened binaries.  While we are unable to produce enough wide binaries to explain their abundance, the orbital properties of ones we do produce are consistent with known wide binaries.
\end{abstract}

\section{Introduction}

An appreciable fraction of Trans-Neptunian Objects (TNOs) are in binary systems, with their quantity and variety depending greatly on which TNO population they inhabit.  This binary fraction ranges from at least 30\% in the low inclination Cold Classical Belt to at least 10\% in the higher inclination Hot Classical Belt \citep{Noll_2008,Fraser_2017}.  The systems that this paper considers are those that have exceptionally wide orbital separations.  Ultra-Wide Trans-Neptunian Binaries (Ultra-Wide TNBs) are some of the more extreme objects in the solar system.  Their individual components are typically of near equal size and have orbital separations far above those of most Trans-Neptunian binaries (TNBs) \citep{Petit_2011}.  2001 QW$_{322}$ is the binary with the widest known separation of approximately 102,000 km, despite neither of the pair being larger than 130 km in diameter \citep{QW_Disc, Petit432, GRUNDY201962}.  Though 2001 QW$_{322}$ is the most widely separated of the known Ultra-Wide TNBs, several other systems are known to have similar extremity \citep{Parker_2011}.  

The incredible separation of these binaries stands out in a Kuiper Belt that still provides them many close encounters with other bodies.  The inaugural work on this topic, \citet{PETIT2004409}, studied three mechanisms in which an Ultra-Wide TNB may become dissociated and each mechanism's associated timescale; these being disassociation through collisions with passing bodies, disassociation through gravitational perturbation of passing bodies, and the complete destruction of one of the binary components to an especially energetic collision.  \citet{PETIT2004409} concluded that by a large margin, non-destructive collisional disassociation was the most relevant in the evolution of TNBs.  

Following this, later studies into this process have generally accounted for collisional interactions with the wider Kuiper Belt exclusively.  They have shown that all known wide binaries can survive into the modern day given reasonable assumptions of the Size Frequency Distribution (SFD) of small TNOs \citep{Parker_2011b,Nesvorn_2021}.

On formation, the work of \citet{Nesvorn__2010} has shown that Ultra-Wide TNBs of near equal mass can be formed very efficiently in the primordial Kuiper Belt due to gravitational instability \citep{Youdin_2005} from excess angular momentum in collapsing clouds.  However, the idea that Ultra-Wide TNBs can form more recently in the modern Kuiper Belt has also been explored by \citet{Parker_2011b} who have shown that such formation is possible though the mutual orbital properties of such binaries tend to evolve away from observed distributions.

Other studies have looked into the effects of Kozai oscillations and tidal friction \citep{kozai1962secular, Perets_2009, Porter_2012, Brunini}.  In this context, Kozai Oscillations are the periodic changes in a system's eccentricity and inclination caused by an outside perturbing force which in this case is the Sun.  These oscillations preserve a binary's semi major axis and the value $cos(I_b) \sqrt{1-e^2_b}$ where $I_b$ is the binary system's mutual inclination relative to the Sun and $e_b$ is its mutual eccentricity.  Tidal friction here has the effect of circularizing an eccentric orbit.  Thus, as detailed by \citet{Brunini}, because high inclination can become high eccentricity (and vice versa) and eccentricity lowers over time, the inclination of wide TNBs can decrease over time; potentially leaving a lasting mark on the mutual inclination distribution of modern day Kuiper Belt binaries.  

Most prior studies, though, in accordance with the results of \citet{PETIT2004409} looked into the problem of Ultra-Wide TNB formation and survival from the perspective of collisional evolution.  Though a landmark work, their study into the most significant driving fore behind binary loss assumes that these forces manifest as a single event, such as a single collision or a single gravitational encounter.  Here, we study the oft-ignored process of gravitational perturbations on these binaries given multiple encounters over time.  We also show that these interactions are a possible mechanism for the gradual evolution of initially stable and tight TNBs into Ultra-Wide TNBs. 

In the next section, we discuss the relevant structure of the Kuiper Belt and which portions of the belt are most relevant in the evolution of Ultra-Wide TNBs.  Section 3 describes our model for gravitational interactions and our initial binary parameters; the results of which are outlined in Sections 4 and 5.  The last section is devoted to our conclusions. 

\section{The Trans Neptunian Region}

We  divide the Kuiper Belt along the same lines as \citet{Gladman2008NomenclatureIT} do; dividing it into three populations based on the orbital properties of the bodies in them.  \par

\begin{enumerate}
  \item Resonant Population:  These objects occupy mean motion orbital resonances with Neptune.  
  \item Scattering Population:  These objects are likely to have one or more close encounters with Neptune on relatively short (10 Myr) time scales which would result in a large change in their semimajor axis.
  \item Classical Population or Main Classical Belt (MCB):  These objects have perihelia distant enough from Neptune that their orbits are stable over relatively long timescales.  
\end{enumerate}

The classical belt in particular can be divided further with respect to solar inclination distribution.  This bimodal distribution is composed of the dynamically Hot main Classical Belt (HCB) and the dynamically Cold main Classical Belt (CCB) in overlapping portions of the Kuiper Belt \citep{Gladman2008NomenclatureIT, Gulbis_2010}.

\begin{enumerate}[3.a]
    \item Cold Main Classical Belt (CCB):  These objects have a narrow solar inclination width of around 5$^{\circ}$.  Ultra-Wide TNBs tend to be found here \citep{Noll_2008, Parker_2011}.
    \item Hot Main Classical Belt (HCB):  These objects have a much wider solar inclination distribution reaching inclinations as high as 35$^{\circ}$ \citep{Noll_2008}.
\end{enumerate}

Though the foundational difference between these two populations is solar inclination, CCB bodies also tend to have a higher albedo and a redder color than HCB bodies.  These differences as well as their differing SFDs indicate that these two populations likely had different formation histories \citep{Petit_2011,Fraser_2014}.
\par

Most relevant to this study are the different properties of each population's binary systems.  Much like most of the rest of the solar system, binaries in the HCB are generally composed of a single massive body with a much smaller companion orbiting it.  But in the CCB, $\sim30\%$ of observed bodies are in binary systems whose component size ratios tend towards parity, and whose separations tend to be wider than the HCB \citep{Noll_2008,NOLL2020}.  Additionally, the aforementioned tendency for CCB bodies to be red does not hold for wider binaries \citep{Fraser_2017,Fraser_2021}.  Binary separation is often expressed in terms of percentage of a binary's Hill radius (R\textsubscript{H}) whose equation is defined as:

\begin{equation}
    r_{hill} = a_{helio}(1-e_{helio}) \sqrt[3]{ \frac{m_{binary}}{3 M_\odot}}
\end{equation}

Beyond a 0.07 R\textsubscript{H} separation, a binary is considered Ultra-Wide and nearly every one known is in the CCB \citep{Noll_2008, Parker_2011}.

\subsection{Relevant TNO Populations}

In this work we consider the influence that close encounters with other TNOs have on the dynamical evolution of Ultra-Wide TNBs. In doing this, we determine which Kuiper Belt populations have the most influential interactions with CCB binary systems and are thus important to include in our simulations.  However, the priorities for determining which populations have the most influence in purely gravitational encounters are different than those of collisional encounters.  For instance, the most significant collisional encounters are those with very high collisional velocities whereas the opposite is true for gravitational encounters.  The significance of a population generally boils down to two criteria.  First, this population must have a high encounter rate with CCB binaries and thus provide many encounters to drive their evolution.  Second, the velocities of these encounters must be low for the effects of the perturbing gravity to be maximized.

In judging a Kuiper Belt population on the basis of these two criteria, we turn to the work of \citet{Abedin_2021}. They have determined both the encounter rate and average encounter velocity of several Kuiper Belt populations with the CCB.

The first population to consider is the Resonant Population. \citet{Abedin_2021} examines three different resonant populations, inner (4:3, 3:2), main (5:3, 7:4), and outer (2:1, 7:3, 5:2).  Of these three, the inner and main resonances provide a mixture of high encounter rates and low encounter velocities. The outer population however has almost half the encounter rate as either of the other two.

\par
The Scattering Population is composed of bodies with large eccentricities and perihelia near to Neptune's orbit \citep{Gladman2008NomenclatureIT}.  These bodies have had a close encounter with Neptune which put them into their current orbit, and likely will do so again in the future.  Due to their wide orbit and high eccentricity, bodies from the Scattering Population do not provide a significant portion of encounters with CCB binaries; failing our first criteria.  Additionally, any scattering bodies that by chance encounter a CCB binary will likely have very high relative velocities \citep{Abedin_2021}; failing our second criteria.  \par

From the Classical Population, the self-interaction of the CCB is the most influential component of our study.  In addition to the average velocity between encountering bodies being very low, the encounter rate is very high \citep{Abedin_2021}.  This is simply because CCB bodies share both the same location in the Kuiper Belt and the same mild orbital properties.  The interaction of the HCB as well is significant for similar reasons.  Though because of its more extended inclination distribution, its encounter rate is lower than that of the CCB self-interaction and its encounter velocity is a bit higher \citep{Abedin_2021}.  Nevertheless, the many encounters stemming from this interaction are slow enough for us to consider it significant.  
\par

In all, the Kuiper Belt populations whose encounters with the CCB we simulate are, the HCB, the Inner Resonant Population, the Main Resonant Population, and the CCB itself.

\section{Methods}

In this section, we outline all of the steps taken in setting up our numerical simulations of gravitational encounters between TNBs and passing TNOs.  A number of specific Kuiper Belt parameters must be determined or assumed.  We need to find the total number of encounters a CCB binary would experience over 4 billion years, the relative velocity of these encounters, and the masses of these passing TNOs.  

\subsection{Encounter Rates and Velocities}

The two parameters we seek first are the encounter rates (mean intrinsic probabilities of encounter) of the CCB with every other population we discussed in the previous section, and the velocities of such encounters. For this, we turn back to the work of \citet{Abedin_2021}.  They have very helpfully determined just these things for each population with the CCB. 

However, we wish to sample from a velocity distribution, rather than to chose a single velocity value to assign for each encountering body.  To this end, we assume that the encounter velocity distribution is a Boltzmann distribution whose mean, we assign to be the average velocity indicated in \citet{Abedin_2021}.  By assuming a Boltzmann distribution, we are still able reproduce the fifth and ninety fifth percentile velocities which they give.

Our velocity distributions are plotted in Figure 1.  The only further modification we make is setting a lower velocity limit at 20 m/s.  Though incredibly rare (0.002\% of all encounters), these events can cause integration errors as strong gravitational focusing leads to encounter timescales unresolvable with our integration step size.  Any encounter speeds which would fall below 20 m/s are simply set to 20 m/s.

\begin{figure}[h!]
\centering
\includegraphics[scale=0.8]{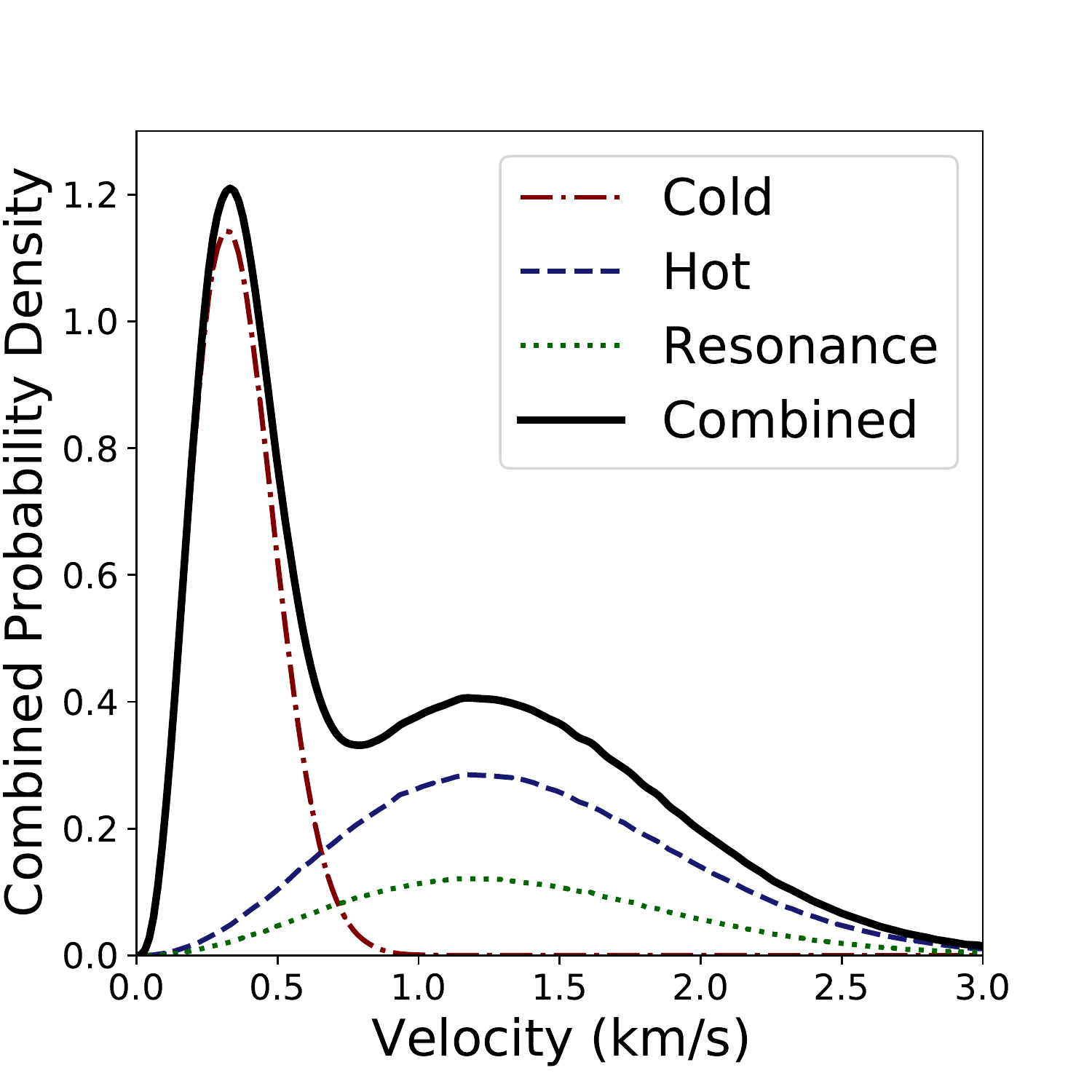}
\caption{The combined probability density of encounter speeds for a CCB binary from each TNO population.  Also plotted are the individual component population speeds with the resonant populations merged into one.}
\end{figure}

\subsection{TNO Size Distribution}

The masses of encountering TNOs and the number of encounters a binary experiences depends on the TNO Size Frequency Distribution (SFD).  For the dynamically hot populations (The HCB, and Inner and Main resonance), we utilize the ``Knee" model distribution as described by \cite{Lawler_2018}.  This is an absolute magnitude ($H$) distribution that consists of two power laws joined together at a sharp but continuous division, dubbed, the knee, or the break magnitude ($H_b$).  The power law at either end of the knee is expressed by 
$\frac{dN}{dH} \propto 10^{\alpha H}$, equivalently expressed in diameter space by 
$ \frac{dN}{dD} \propto D^{-(5 \alpha + 1)} $
.  The power law governing smaller bodies (the faint end) is defined with $\alpha_f = 0.4$ and likewise, the law governing larger bodies (the bright end) is defined with $\alpha_b = 0.9$.  While \citet{Lawler_2018} are able to fit the $H$ distribution well with a knee power law, due to observational uncertainty, the exact break magnitude is somewhat unconstrained.  To probe the possible parameter space, we use break magnitudes of 9.1 and 7.9. 
\par

The $H$ distribution of the CCB, however, follows a different curve.  We thus turn to \citet{Kavelaars_2021}, who have determined the distribution to be an exponentially tapered power law.  We utilize this SFD along with their fit fixed faint power law slope of $\alpha_C = 0.4$. 

\par
Of course, a magnitude distribution does not translate directly into size.  We do need to know or at least assume the albedo of the bodies we are modeling, though only a small minority of TNOs have had their albedos measured.  For the dynamically hot population (the HCB and both resonance populations), we use an albedo value of 0.04 as this is the value used by \citet{Lawler_2018} when they determine their split power law distribution.  With this albedo assumption, we can translate our two break magnitudes into radii of 50 km and 85 km respectively.  However, the albedo of the CCB appears to be greater than that of other populations.  With this in mind, we assume a CCB albedo of 0.14 as determined by \citet{Vil2014}.  Additionally, albedo has shown in practice to be a very influential parameter in binary survival lifetimes so we also employ a CCB albedo of 0.12 as a lower extreme.  In all populations, we assume a density of 1.0 g cm$^{-3}$ as this is a value used by \citet{Parker_2011b} and \citet{Nesvorn_2021}.
\par

The SFD that we sample from for our CCB tapers off at low $H$ such that no bodies of $H < 4$ ($r \gtrsim 300$ km) are generated.  This is agreeable with present day observations \citep{Nesvorn_2021} setting a natural upper limit on the size of bodies generated in our simulation for this encounter type.  For our HCB and 3:2 SFD, we set an artificial $H$ limit of 4 ($r = 500$ km) as well.  Each of our SFDs and minimum magnitude limits are set up with the modern Kuiper Belt in mind.  Certainly in earlier epochs, the Kuiper Belt's populations may have had different SFDs to today which we do not account for.
\par

By varying our CCB albedo and our dynamically hot populations' break radius, we thus have 4 total Kuiper Belt combinations that we simulate.

\par

\subsection{Minimum Radius Cutoff}

A consequence of using a mass distribution that is composed of one or two power laws is that as one models smaller and smaller bodies, the number of encounters approaches infinity.  However, the effects from such low mass encounters diminish despite their increased frequency as their mass becomes increasingly small.  Thus, we need to choose a minimum TNO size below which, we can ignore encounters without significantly affecting their aggregate effects in our simulation.  \par

For each of our different SFD shapes, we simulate 5 million encounters between each of our previously established Kuiper Belt models and a stationary body.  These shapes are the CCB tapered power law and the dynamically hot populations' split power law with break radii of 50 km and 85 km.   Encounter direction is sampled from a uniform sphere, reflecting random directions in 3D space \citep{Henon_1972}. Though the masses of these passing TNOs reflect our established distributions, their velocities are fixed.  For each encounter, we record the gravitational impulse imparted on the stationary body in three dimensions.  Smaller bodies, though significantly more numerous, tend to evenly distribute their impulse in three dimensional space, canceling each other out. They do not contribute much to the overall impulse outside of a random walk that is dwarfed by encounters with larger bodies.
\par

Each encounter imparts an impulse onto the central body which is recorded.  To find the smallest radius of encountering bodies that contribute substantially to the evolution of a binary, we pair the radius of every encountering body with the impulse it provides.  The total three dimensional impulse imparted by all bodies in the simulation is recorded and compared to the individual impulses provided by each body.  We arrange the encountering bodies by radius and incrementally add the impulse they provide until we reach 10\% of the total impulse.  The radius where this boundary is reached is the smallest significant radius.  We run 300 such simulations with the lower 5th percentile of results being our calculated minimum significant radius.

Of course, in order to begin drawing from these distributions, we need to assume a minimum radius to draw from or else we run into the same problem described at the beginning of this subsection.  The minimum radius we choose is 5 km.  This radius is appropriate as after some experimentation, using a lower radius does not influence the results of our minimum radius simulations.  
\par

Our resulting minimum significant radius from CCB self interaction is 36.3 km.  From our dynamically hot populations, it varied significantly based on our break radius.  For break radii of 50 km and 85 km, our minimum encountering body radius is 34.0 km and 46.7 km respectively.  Out of an abundance of caution, however, we elect to use a minimum significant encounter radius of 20 km for all of our SFDs.
\par

\subsection{Total Number of Encounters}

Our simulation takes place over billions of years.  The total number of encounters our binary system will have over that time can be calculated with.

\begin{equation}
    N_E = \langle P_i \rangle N R^{2}t
\end{equation}

Where $N_E$ is the number of encounters, $\langle P_i \rangle$ is the mean intrinsic probability of encounter, $N$ is the total number of bodies that may possibly be encountered, $R$ is the maximum impact parameter over which we consider encounters, and $t$ is the time over which an encounter can happen.  Our intrinsic probability of encounter parameter is discussed in Section 3.1 but we have not yet talked about $N$ and $R$. \par

In order to determine the total number of bodies that may be encountered by a CCB binary, we need a definitive number of known TNOs above a certain size to which we can anchor our SFDs.  Each of our populations have their own anchor.  For the HCB, we turn to \citet{Petit_2011}, who estimates the number of bodies brighter than $H < 8$ to be 4,100.  For our various resonant populations, we turn to \citet{Gladman_2012} who estimate the number of Inner Resonance and Main Resonance bodies brighter than $H < 8$ to be 1270 and 750 respectively.  These values anchor and calibrate our SFDs, allowing us to generate the proper number of potential encounters down to an arbitrary minimum radius.  The SFD of the CCB, already defined with an anchor, does not need further calibration.  As for $R$, the maximum impact parameter, we use a value of $3.6 \times 10^6$ km, or 8 times the R\textsubscript{H} of 2001 QW$_{322}$

\par

However, such calculations produce a large amount of encounters that are not especially significant.  Passages of small TNOs that never get much closer than 4 R\textsubscript{H} dominate such encounter generators.  Thus we discard TNOs whose mass is not large enough to have a significant impact on our binary system during its flyby.  The space around our binary is divided into 3 zones depicted in Figure 2.  The first encompasses the inner 1 R\textsubscript{H}, the second encompasses the space between 1 and 4 R\textsubscript{H} and the last is the space between 4 and 8 R\textsubscript{H}.  TNOs whose encounter distance is further than 4 R\textsubscript{H} would have to be very massive indeed to affect a binary at all, so only passing TNOs of mass greater than $1.1 \times 10^{20}$ kg ($R$ $\gtrapprox$ 300 km) are simulated; all others are discarded.  In the second zone, only TNOs of mass greater than $2.6 \times 10^{17}$ kg ($R$ $\gtrapprox$ 40 km) are simulated.  In the innermost zone, all encounters are simulated.  With some testing, we find that simulating more expansive coverage of low mass encountering bodies has little impact on the results of a simulation.

\begin{figure}[h!]
\centering
\includegraphics[scale=0.8]{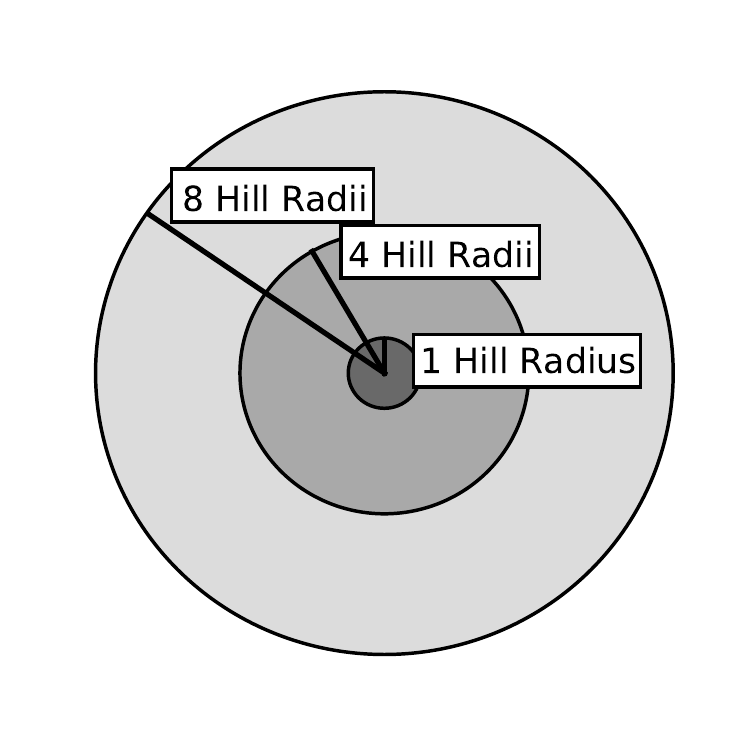}
\caption{The three zones around the binary system.  Bodies with an encounter separation between 4 and 8 R\textsubscript{H} must have more mass than 1.1 $\times 10^{20}$ kg.  Bodies with an encounter separation between 4 and 1 R\textsubscript{H} must have more mass than $2.6 \times 10^{17}$ kg.  Bodies of any mass can pass closer than 1 R\textsubscript{H}.}
\end{figure}

\subsection{Simulating Encounters}

We use a modified version of the SWIFT RMVS 4 integrator to simulate the passing TNO encounters \citep{1994Icar..108...18L}.  In this modified scenario, the primary companion is given system mass corresponding to the binary we are simulating and sits at the origin of our coordinate system.  Our secondary binary companions are massless and bear the orbital properties of the systems we simulate.  While their initial semi major axes, eccentricities, and inclinations will be set specifically for each type of binary we simulate, their arguments of perihelion, longitudes of ascending node and mean anomalies are always randomized.  Our simulation time step is always at most 1/20 the orbital period of the binary we simulate; and the influence of the Sun is accounted for.  Our Sun is drifted on a circular orbit 44 AU away corresponding to the heliocentric semi major axis of the typical CCB binary.

At their prescribed time, interacting TNOs are introduced to the simulation 8 R\textsubscript{H} distant with velocities consistent with our established distributions from Section 3.2 and random trajectories.  After they pass through the system and reach 8 R\textsubscript{H} again, they are removed.  However, if the incoming TNO has a high velocity and thus spends less time perturbing the binary, we do not simulate its passage.  We instead use the impulse approximation to perturb the primary and its swarm of test binary companions.  The velocity of an incoming TNO is considered ``high" when the encounter timescale (or closest approach over encounter velocity) is less than 50 days, or 10 times the shortest integration time step we ever use.

Our test bodies, if perturbed beyond 8 R\textsubscript{H} of the primary companion, are assumed to have been unbound and are removed from the simulation.  Additionally, if the separation between a test body and central mass is ever less than the sum of their physical radii, they are assumed to have collided and the test body is similarly removed.  At the end of their prescribed simulation time, the simulation is ceased.

\subsection{Fast Simulations}

Ideally, each of our simulations would fully account for every single year of evolution with each encounter time randomly distributed in that time span.  However, in the interests of building a larger statistical sample of simulated systems, we can compress our simulated time without sacrificing the validity of our results.

In the simulation configuration described thus far, the amount of time that an interacting TNO is present in the system is very short compared to the overall simulation time where no encounters are happening.  Were it not for the exterior gravitational force provided by the Sun, the test bodies would move in perfectly Keplarian orbits in this time between encounters.  By not simulating all of this non-encounter time, our simulations can be much faster and require fewer resources to run.  

This leaves only the Sun to be considered. Its gravity is the only force acting on the binary between encounters so in compressing this time, we must also simulate the Sun's influence on the system.   In practice, the vast majority of the Sun's influence manifests in the form of Kozai cycles.  While this does not have an effect on the mutual semimajor axes of the binary systems, it can cause their eccentricities and inclinations to oscillate \citep{kozai1962secular}.  One significant consequence of this is that the two binary components may collide once their eccentricity becomes too high and their periapsis becomes too low.  These collisions very rarely happen in simulations that do not include the Sun.  Thus, our metric for accurately accounting for billions years of solar influence, is comparing the number of binary collisions in them, with those in simulations which fully account for billions years of solar influence. 

Through a series of trial runs, we empirically find that cutting inter-encounter time from a typical value of 700,000 days to 12,000 days still captures an adequate sampling of solar perturbations; since collisions between both binary components happen just as often as in the uncompressed simulation runs.  Because Kozai cycles are cyclical and in our case, operate on timescales of tens of thousands of years, we do not need to simulate the entire billions of years of solar influence to account for them.  This modest delay between encounters allows our simulations to run up to 50 times faster, while fully reproducing the number of binary collisions that we see otherwise.  As such, all simulations in this paper utilize this compression, though data and graphs are depicted assuming uncompressed time.

\section{Results}

\subsection{Significance of Iterative Perturbative Encounters}

With all of our parameters in place, we reevaluate and test the effect of gravitational perturbations from TNO flybys on a given CCB binary system.  Simulating these interactions on a binary that is allowed to evolve, we find that there are two primary mechanisms of binary loss.

The first of these binary loss mechanisms is collision between the two components of the binary themselves.  As briefly described in Section 3.7, this is caused primarily by Kozai oscillations driven by the Sun that dramatically raise the eccentricity of a mutually inclined binary system.  Once this eccentricity approaches 1, the binary periapsis drops low enough that both components collide.  Even initially uninclined binaries may succumb to this type of disassociation as their inclinations can be altered by strong passages.  This collision is not the result of any one TNO passage and with an initial inclination close enough to 90\textdegree, can occur without any passages.  It should be noted, however, that our model does not include the effects of tidal friction which prior work has found can tighten and circularize the orbits of eccentric binaries \citep{Perets_2009, Porter_2012, Brunini}.  Thus, it is possible that destruction due to Kozai induced collisions may often be unphysical and simply a consequence of neglecting this effect.  

The second mechanism of binary loss is ejection, or disassociation resulting from a binary's gravitational binding energy being exceeded by the tidal potential of a passing TNO.  This mechanism of binary loss is the focus of the analytical calculation done by \citet{PETIT2004409}.  However, we find such binary loss is often due to many encounters slowly widening the separation over time.  This gradual widening of a binary's separation is very effective at making it more fragile.  Eventually, the binary becomes fragile enough for a single encounter to break it.  

When testing the long term stability of specific Ultra-Wide binaries, this increasing separation proved to be a very influential.  This is shown in Table 1 where in any Kuiper Belt SFD, between 72 and 82 percent of all lost binaries whose initial parameters are that of 2001 QW$_{322}$ become lost only after experiencing prior orbital widening of at least 4.5\% R\textsubscript{H} or 20\% of their initial separation.

\begin{table}
\begin{tabular}{ |ccc|  }
\hline
 CCB Albedo & $R_b = 50$ km & $R_b = 85$ km \\
 \hline
 \hline
 $\alpha = 0.12$ & 0.82& 0.77\\\cline{1-1}
 $\alpha = 0.14$ & 0.77& 0.72\\
 \hline
\end{tabular}
\caption{The fraction of lost 2001 QW$_{322}$-like binaries that become dissociated following an outward migration of at least 4.5\% R\textsubscript{H}.  All such binaries begin with a separation of 22\% R\textsubscript{H}.  $\alpha$ refers to our assumed CCB albedo and $R_b$ refers to the break radius of our split power law dynamically hot populations.}
\end{table}

The first of these loss mechanisms, collision between the two binary components, is far less likely to occur in simulations that do not account for Kozai mechanisms.  The second mechanism, the ejection of a binary component, is a process greatly dependant on the gradual widening of a binary's separation.  This gradual widening and eventual loss would not be included in \citet{PETIT2004409}, possibly dulling an estimated impact of gravitational erosion.  \par

\subsection{Wide Binary Lifetimes}

The longevity of certain wide TNBs has been tested in the past assuming a collisionally active Kuiper Belt \citep{Parker_2011b, Nesvorn_2021}.  To compare their results to our gravitationally active Kuiper Belt model, we also test this longevity.  With these results, we can determine just how long one might expect these binaries to survive in the environment they now inhabit.  Under the effects of many perturbative encounters with KBO bodies, binaries are disassociated at a rate greatly resembling that of exponential decay, being represented by,

\begin{equation}
    N(t) = N_oe^{-t \, (a/\tau) }
\end{equation}

Where $\tau$ is the mean lifetime of the binary system. 

In our study, we have simulated three specific known binaries to determine how long they may remain in the Kuiper Belt without disassociating.  The first is 2001 QW$_{322}$ ($22.2\%$ R\textsubscript{H} separation) as it is the widest known TNB and has been looked at specifically by \citet{Parker_2011b}.  Our second is 2000 CF$_{105}$ ($16.8\%$ R\textsubscript{H} separation) as in that same paper, it has shown to be the binary most sensitive to collisions.  Our third is 2004 JZ$_{81}$, representing an Ultra-Wide TNB with a comparatively low separation ($9\%$ R\textsubscript{H} separation) \citep{Parker_2011b}.  The decay rate of 2000 CF$_{105}$ under each Kuiper Belt SFD are shown in Figure 3.  Each of these three binaries are subject to 10 billion years of Kuiper Belt interactions.

\begin{figure}[h!]
\centering
\includegraphics[scale=0.8]{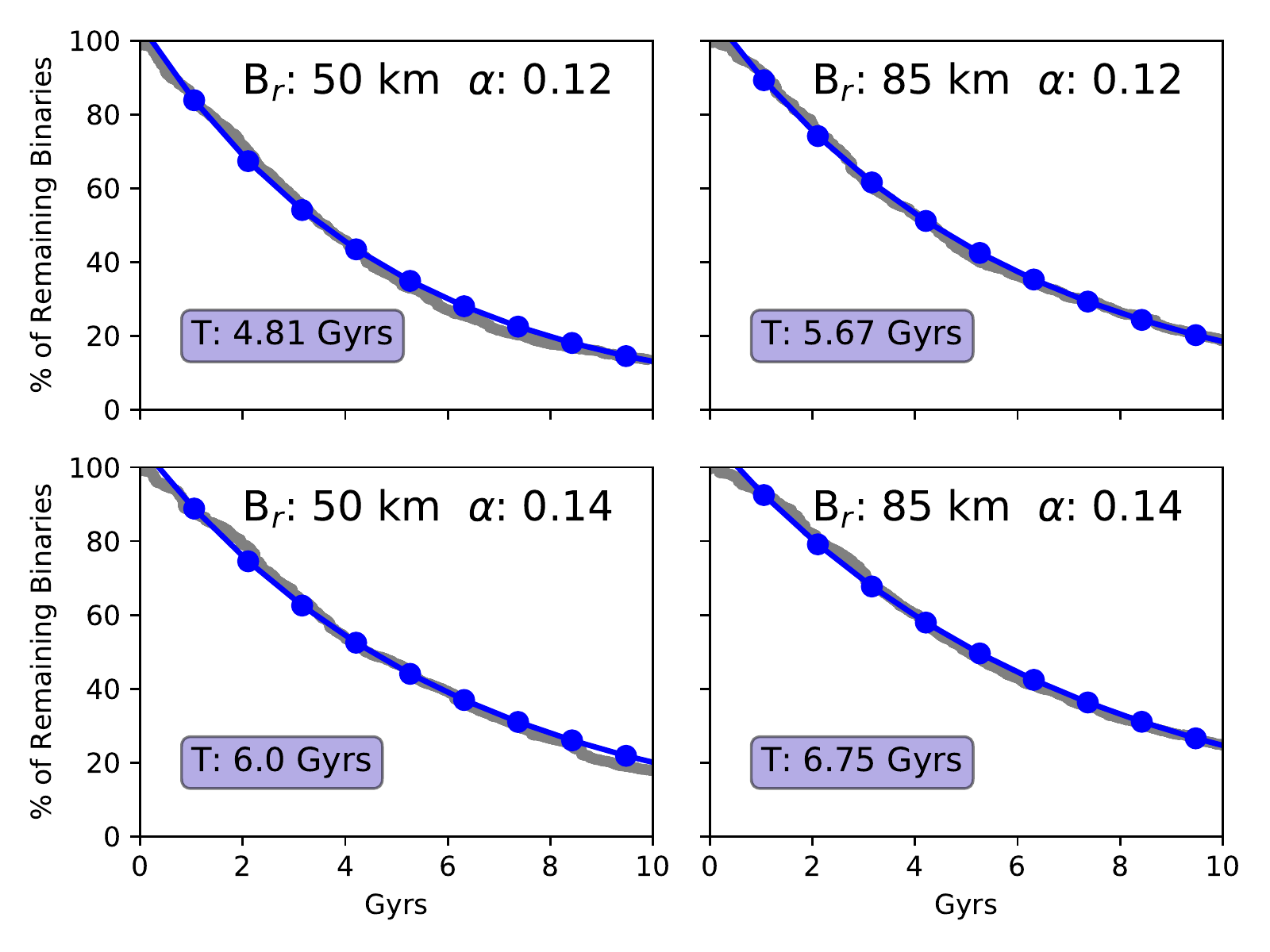}
\caption{The dark gray points are the percentage of surviving binaries over time whose initial separation is that of 2000 CF$_{105}$ starting at 100.  The blue dotted line represents a fit exponential decay curve.  Each panel represents the different Kuiper Belt parameters simulated and includes a box showing the average binary lifetime in that SFD.  B$_r$ is break radius of the dynamically hot populations and $\alpha$ is albedo of the CCB.}
\end{figure}

If we assume a CCB albedo of 0.14, then a binary like 2000 CF$_{105}$ would be expected to survive an average of 6-7 billion years in the modern Kuiper Belt before disassociating.  Thus, if such binaries were formed in the early Kuiper Belt, we would expect 50-55\% of an initial population of 2000 CF$_{105}$ like binaries to survive 4 billion years under the influence of gravitational encounters alone, given what we know of the modern Kuiper Belt.  The exact average lifetimes of each binary under each Kuiper Belt SFD are listed in Table 2.

\begin{table}
\begin{tabular}{ |cccc|  }
\hline
 TNO Binary & CCB Albedo & $R_b = 50$ km & $R_b = 85$ km \\
 \hline
 \hline
 \multirow{2}{*}{2001 QW$_{322}$}& $\alpha = 0.12$ & 5.81& 6.15\\\cline{2-2}
 & $\alpha = 0.14$ & 6.60& 7.43\\
 \cline{2-4}
 \multirow{2}{*}{2000 CF$_{105}$}& $\alpha = 0.12$ & 4.81 & 5.67\\\cline{2-2}
 & $\alpha = 0.14$ & 6.0& 6.75\\
 \cline{2-4}
 \multirow{2}{*}{2006 JZ$_{81}$}& $\alpha = 0.12$ & 14.4 & 15.0\\\cline{2-2}
 & $\alpha = 0.14$ & 15.1& 19.2\\
 \hline
\end{tabular}
\caption{Our simulated average lifetime of 3 Kuiper Belt binaries in Gyrs under our 4 different Kuiper Belt SFDs.  B$_r$ is break radius of the dynamically hot populations and $\alpha$ is albedo of the CCB.}
\end{table}

These figures imply that given an initial population of even the most delicate Ultra-Wide TNBs, a significant fraction could survive 4 billion years in the modern Kuiper Belt.  However, as they are, these figures are underestimates.  Calculating the initial population of wide binaries by simply examining their average lifetimes ignores a crucial aspect of our simulations.  The mutual separation of a binary can change.  

In our simulations, mutual binary separation usually increases over time but this need not be the case.  2000 CF$_{105}$-like binaries that are subject to inward migration are over-represented among our surviving sample, simply because binaries with lower separations are harder to break.  But if an initially 2000 CF$_{105}$-like binary survives 4 billion years by significantly migrating inward, it can hardly be considered a 2000 CF$_{105}$-like binary anymore.  Thus, to gain an accurate ratio of initial to final populations of Ultra-Wide TNBs, we must consider binaries that survive by migrating inward as lost.

If we allow these binaries to evolve over 4 billion years and introduce a criterion that any surviving binary's final separation must not have decreased by more than 20\% of its initial separation, then our initial binary populations must be higher.  Under the same mass density and CCB SFD as before, we calculate that if subjected to 4 billion years of gravitational perturbations from the modern Kuiper Belt, a 2000 CF$_{105}$-like binary has a likelihood of between 37\% and 44\% of remaining both intact and of comparable separation to their origin.  This assumes the albedo of the CCB is 0.14 and a break radius of 50 km and 85 km respectively.  Table 3 lists the fraction of our previous three binaries that survive 4 billion years under our different Kuiper Belt SFDs.

\begin{table}
\begin{tabular}{ |cccc|  }
\hline
 TNO Binary & CCB Albedo & $R_b = 50$ km & $R_b = 85$ km \\
 \hline
 \hline
 \multirow{2}{*}{2001 QW$_{322}$}& $\alpha = 0.12$ & 0.38& 0.40\\\cline{2-2}
 & $\alpha = 0.14$ & 0.43& 0.45\\
 \cline{2-4}
 \multirow{2}{*}{2000 CF$_{105}$}& $\alpha = 0.12$ & 0.30 & 0.34\\\cline{2-2}
 & $\alpha = 0.14$ & 0.37& 0.44\\
 \cline{2-4}
 \multirow{2}{*}{2006 JZ$_{81}$}& $\alpha = 0.12$ & 0.79 & 0.80\\\cline{2-2}
 & $\alpha = 0.14$ & 0.81& 0.84\\
 \hline
\end{tabular}
\caption{Fraction of binaries that both survive 4 billion years and do not migrate inward by more than 20\% of their initial separation.  B$_r$ is break radius of the dynamically hot populations and $\alpha$ is albedo of the CCB.}
\end{table}

\subsection{Binary Widening}

We also seek to test the hypothesis that an initially tight binary system can become Ultra-Wide as a result of gravitational perturbative encounters.  If this is the case, we seek to determine how large a population of widened binaries would be given the present conditions of the Kuiper Belt.  For this, we subject a population of binary systems with separations randomly distributed between 3\% and 5\% R\textsubscript{H} to 4 billion years of gravitational Kuiper Belt interactions.  The eccentricities and inclinations of these binaries are sampled from the known non-Ultra-Wide TNB sample in the modern Kuiper Belt \citep{GRUNDY201962}.  Each of these binary systems are given a total mass equal to 2001 QW$_{322}$.  

What we find is that binaries that have initially tight separations can in fact become Ultra-Wide due to gravitational perturbations.  An example is shown in Figure 4, coming from one of our simulations with a SFD defined by $\alpha_{CCB} = 0.14$ and $R_B = 50$ km.  Here, our binary begins with a separation around 5\% of its Hill radius where it stably orbits until a large perturbation at around t = 0.4 Gyrs.  After this perturbation, the binary maintains a separation comparable to 2001 QW$_{322}$ and 2000 CF$_{105}$, eventually surpassing them both.  This binary then achieves a separation far higher than any observed today following a perturbation at t = 3.6 Gyrs.  This binary's eccentricity and inclination are guided primarily by Kozai oscillations prior to its first major encounter.  Following this, perturbations from passing bodies become the primary driver behind the changes in the binary's orbital properties.

\begin{figure}[h!]
\centering
\includegraphics[scale=0.3]{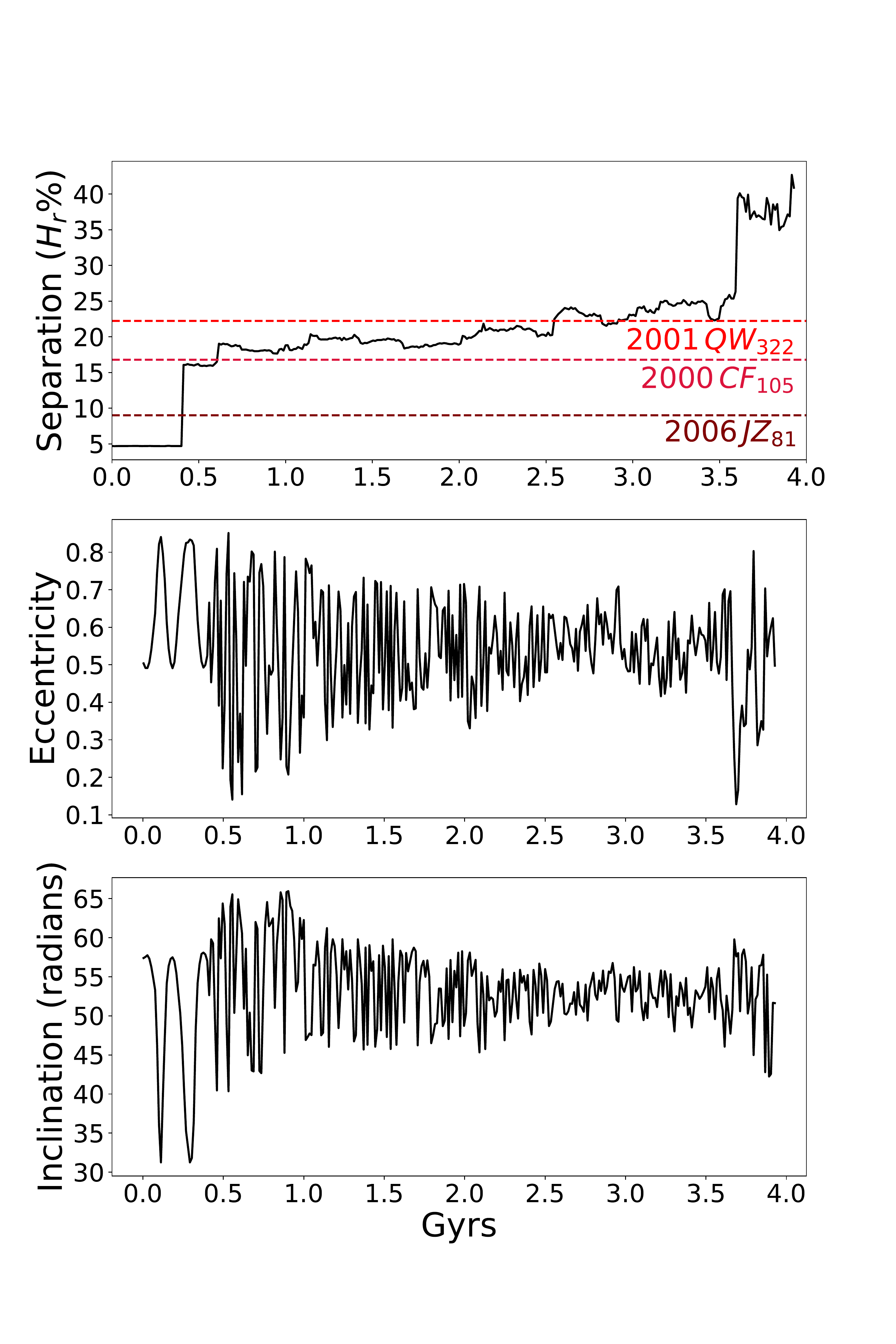}
\caption{Separation (R\textsubscript{H} percentage) vs time (millions of years) of a simulated binary.  This is an example of a binary starting with a separation under $5\%$ R\textsubscript{H} but is eventually widened to beyond that of 2001 QW$_{322}$.  For comparison are  the current separations (in relative Hill radii) of the binaries 2001 QW$_{322}$ ($22\% H_R$), 2000 CF$_{105}$ ($17\% H_R$), and 2006 JZ$_{81}$ ($9\% H_R$).  This wide separation does not last as the binary is lost at around 3.8 Gyrs.  }
\end{figure}

In our example, the binary widens significantly but becomes disassociated before the end of the simulation.  This possibility is to be expected as wide binaries are more susceptible to perturbations than tight ones.  Despite their eventual loss, these binaries can be fairly long-lived.  One could imagine observing a widened binary, noting that is very weakly bound and wondering how such a system could have survived so long.  Thus, instead of measuring the fraction of wide binaries after 4 billion years of simulation, we instead measure the fraction of Ultra-Wide (R $>$ 0.07 R\textsubscript{H}) binaries at any given time between year 3 and 4 billion of our simulation.  This data is shown in Table 4.  Across all assumed CCB albedos and SFDs, we find that $\sim$1\% of all binaries that were initialized at 3-5\% R\textsubscript{H} separations reside in Ultra-Wide configurations billions of years later.

\begin{table}
\begin{tabular}{ |ccc|  }
\hline
 CCB Albedo & $R_b = 50$ km & $R_b = 85$ km \\
 \hline
 \hline
 $\alpha = 0.12$ & 0.015& 0.012\\\cline{1-1}
 $\alpha = 0.14$ & 0.010& 0.009\\
 \hline
\end{tabular}
\caption{Fraction of initial tight (3\% $<$ R $<$ 5\% R\textsubscript{H}) TNBs that are ultra-wide (R $>$ 0.07 R\textsubscript{H}) at any given time.  }
\end{table}

Finally, we note that binaries with starting separation of 1-2\% R\textsubscript{H} are generally immobile.  With no Kuiper Belt SFDs do any significant quantity of binaries ($<$ $.2\%$) become Ultra-Wide; indicating that in the modern Kuiper Belt, these binaries are quite resistant to gravitational perturbations and are not a plausible source for modern Ultra-Wide binaries. 

\section{Discussion}

\subsection{Disassociation Timescales of Known Binaries}

Our testing suggests that in the modern Kuiper Belt, the known Ultra-Wide CCB binaries are generally stable including even the widest binary 2001 QW$_{322}$.  This result agrees with that of \citet{Parker_2011b}, showing that even the inclusion of gravitational perturbative encounters cannot sufficiently disrupt the stability of these binaries assuming reasonable restrictions on the Kuiper Belt's SFD.  The average lifetimes of these binaries can be comparable to the age of the solar system.  For 2000 CF$_{105}$ in particular, assuming $\alpha_{CCB} = 0.14$, its estimated lifespan is between 6 and 7 billion years.  Even accounting for a possible separation change over time, after 4 billion years of perturbations, between 35 and 45\% of binaries as loosely bound as 2000 CF$_{105}$ would remain.

This separation change as well plays an important part in the disassociation rate of binary systems.  A similar effect is observed by \citet{Parker_2011b} in the case of collisional interactions among TNBs.  Such evolution is not accounted for in the original work of \citet{PETIT2004409} which may affect their determination of the relative importance of TNO encounter types in TNB disassociation rates.  

One thing that becomes apparent is that the average lifespan of our tested Ultra-Wide binaries are very dependant on the albedo of the CCB.  We assume an average CCB albedo of 0.14 as this is what is estimated by \citet{Vil2014}.  But the directly measured R-band geometric albedo of 486958 Arrokoth is 0.21 \citep{Hofgartner_2021}, considerably higher than our assumption.  If it should be the case that the average albedo of the CCB is higher than 0.14, the observed Ultra-Wide CCBs are likely to be more stable.  Additionally, our binary's lifetimes seem to be weakly dependant on our assumed break radius of the dynamically hot portion of the Kuiper Belt, with a larger break radius favoring slightly longer binary lifetimes.

However, our results cannot be used to determine the likelihood of any specific binary surviving unscathed since the formation of the Kuiper Belt.  Our models only take into consideration the Kuiper Belt as it currently is observed.  In the past, it had a greater population than present.  Thus, we cannot compare our results with those of \citet{Nesvorn_2021} who considered a dynamically changing Kuiper Belt.  Given the larger number of perturbing bodies in the historical Kuiper Belt, it would seem likely that our estimated binary lifetimes are in fact overestimates.

\subsection{Evolution Towards Ultra-Wide}

While we do show that Ultra-Wide TNBs readily form from a population of initially tight binaries subjected to gravitational perturbations, we do not produce a population size consistent with observations. Even in our most optimistic scenarios, our Ultra-Wide ($a > 7\% R_H$) binary fraction is never greater than 1.6\%.  In contrast, we would expect this ratio to be at least 5\% \citep{Lin_2010}.  The degree of widening of our binaries is shown in Figure 5.  Our population of binaries is sampled in each simulation, every 5 million years from 3-4 Gyrs in our 4 Gyr simulations.  We can reliably produce Ultra-Wide binaries in excess of 14\% Hill radius separation but our quantity of Ultra-Wide binaries consistently falls short of predictions.

\begin{figure}[h!]
\centering
\includegraphics[scale=0.8]{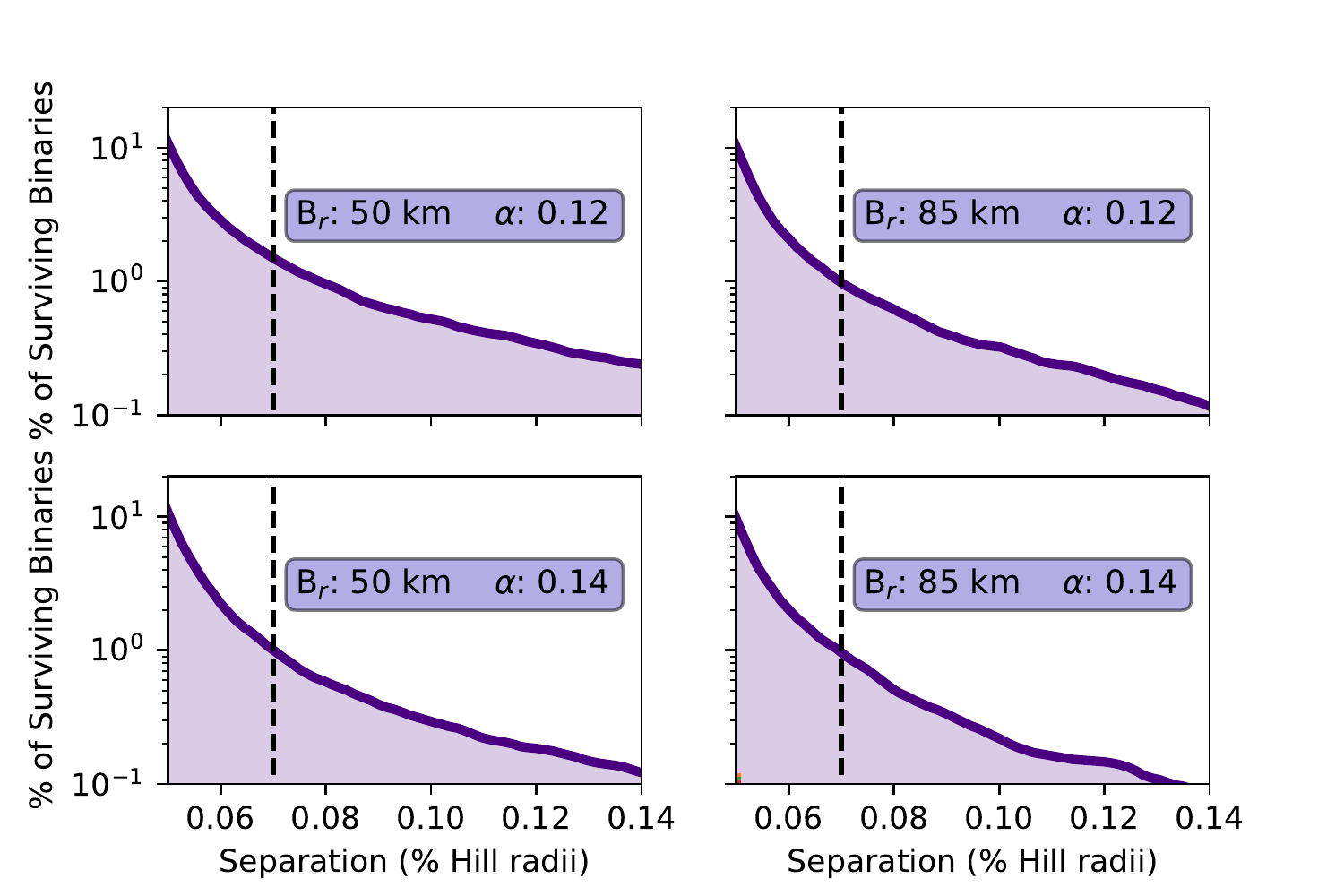}
\caption{A cumulative distribution of the separations of each sampled binary.  A line is drawn at 7\% Hill radius as this marks where a binary becomes Ultra-Wide.  Each panel represents a Kuiper Belt SFD we have modeled; in none of them does the fraction of Ultra-Wide binaries rise over 2\%.}
\end{figure}

Additionally, we begin with a population of binaries that, while not Ultra-Wide, are wider on average than the typical Kuiper Belt binary.  64\% of known Kuiper Belt binaries have less separation than 2\% of their own Hill radius \citep{GRUNDY201962} and these binaries generally do not widen.  The actual fraction of $> 2\%$ R\textsubscript{H} binaries are also likely to be even higher given their increased difficulty of detection.  We also make the assumption that tight precursor binaries have very similar orbital makeups to the tight binaries we currently observe in the Kuiper Belt.  It is of course possible that these precursor binaries have properties unique to an earlier era of the Kuiper Belt and thus are not seen today.  

However, it should be noted again that the models we have used of the Kuiper Belt have not been diachronic.  We have only been simulating the Kuiper Belt as it currently is.  In the more crowded era of the early Kuiper Belt, the larger number of perturbers may magnify this effect and create a larger population of Ultra-Wide binaries.  

\subsection{Properties of Widened Binaries}

Ultra-Wide TNBs have been shown to have different mutual orbital properties than tight (R $<$ 7\% R\textsubscript{H}) binaries.  Their orbital separations are of course higher but they also seem to have different eccentricity and inclination distributions.  Among our simulated binaries that become Ultra-Wide over the course of our simulations, we observe similar orbital properties.  These widened binaries are sampled from 3-4 billion years into our simulation as described in section 5.2.

The mutual eccentricities of Ultra-Wide TNBs tend to on average be larger than those of tighter binaries.  The average eccentricity of Ultra-Wide CCB binaries is 0.53 compared to the average value for tight binaries of 0.34 \citep{GRUNDY201962}.  While a wide binary's eccentricity is generally high, the larger distribution is not thermalized as might be expected in an eccentricity distribution greatly influenced by outside perturbations.

In our simulations, our widened binary eccentricity distribution greatly resembles that of the known Ultra-Wide binary sample as shown in Figure 6.  It remains more extended than the known tight binary eccentricity sample while still not resembling a thermalized distribution.  Figure 6. depicts a widened population from our Kuiper Belt SFD defined by $\alpha_{CCB} = 0.14$ and $R_B = 50$ km but the resemblance holds true with every SFD which we simulated.

\begin{figure}[h!]
\centering
\includegraphics[scale=0.8]{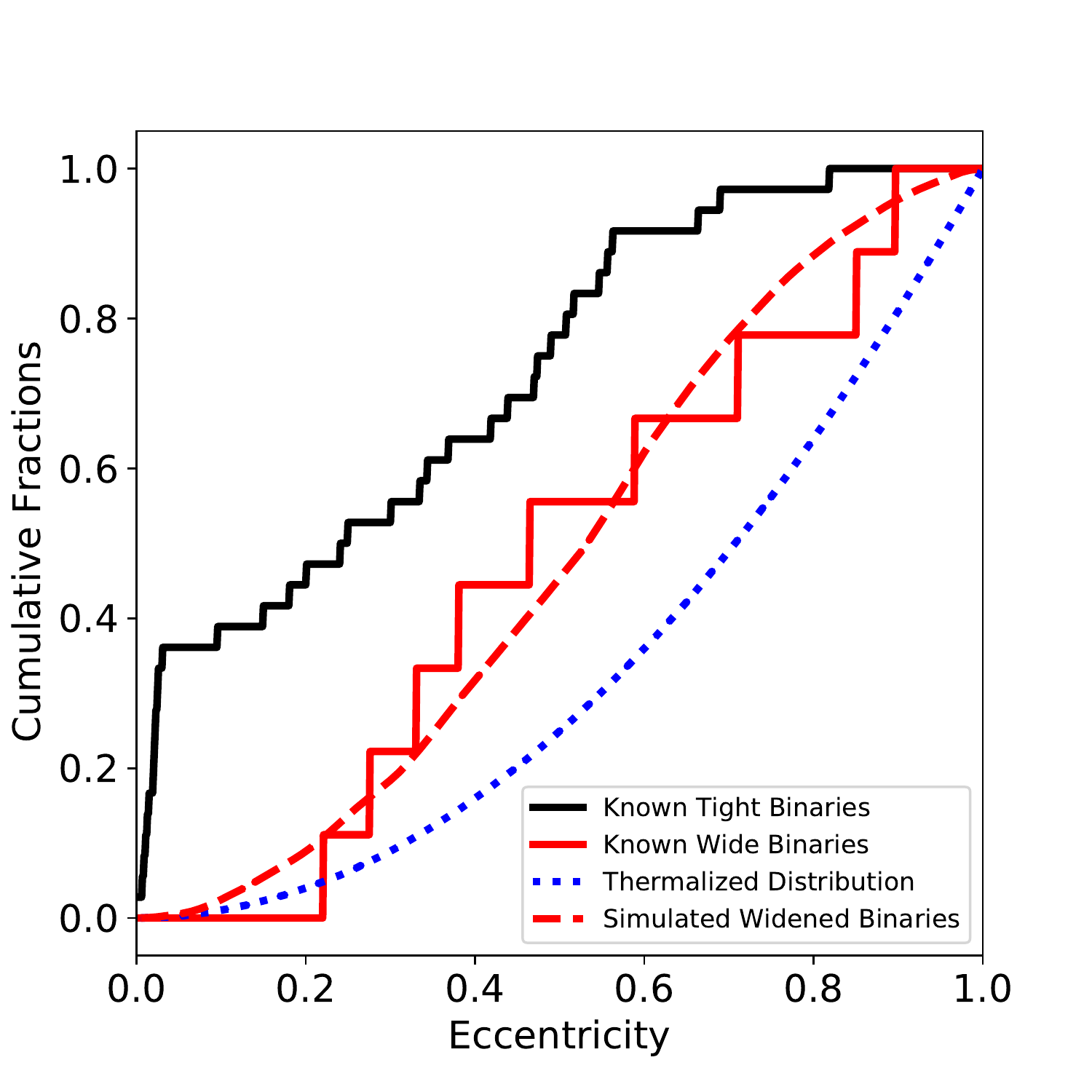}
\caption{The eccentricity distribution of our widened binaries compared to the known distribution of both tight and Ultra-Wide binaries.  Also plotted for comparison is a thermalized eccentricity distribution.}
\end{figure}

In terms of inclination, the differences between the observed tight and Ultra-Wide components of the CCB are more complex.  Tight CCB binaries are very heavily biased towards prograde orbits but have a strong tendency towards higher mutual orbital inclination.  In contrast, a third of all known Ultra-Wide TNBs are retrograde.  In addition, these orbits prefer a more planar inclination orientation, with a dearth \text{of} binaries having mutual inclinations between 55$^{\circ}$ and 125$^{\circ}$ \citep{GRUNDY201962}.  

Among our widened binaries, we are unable to replicate this inclination distribution as well as we do with eccentricity.  As pictured in Figure 7, our binaries begin with an inclination distribution sampled from the known tight binary population and eventually assume a slightly more planar distribution.  What is pictured is again taken from our simulation whose Kuiper Belt SFD is defined by $\alpha_{CCB} = 0.14$ and $R_B = 50$km.  But again, this result is common to all of our SFDs.  

\begin{figure}[h!]
\centering
\includegraphics[scale=1.0]{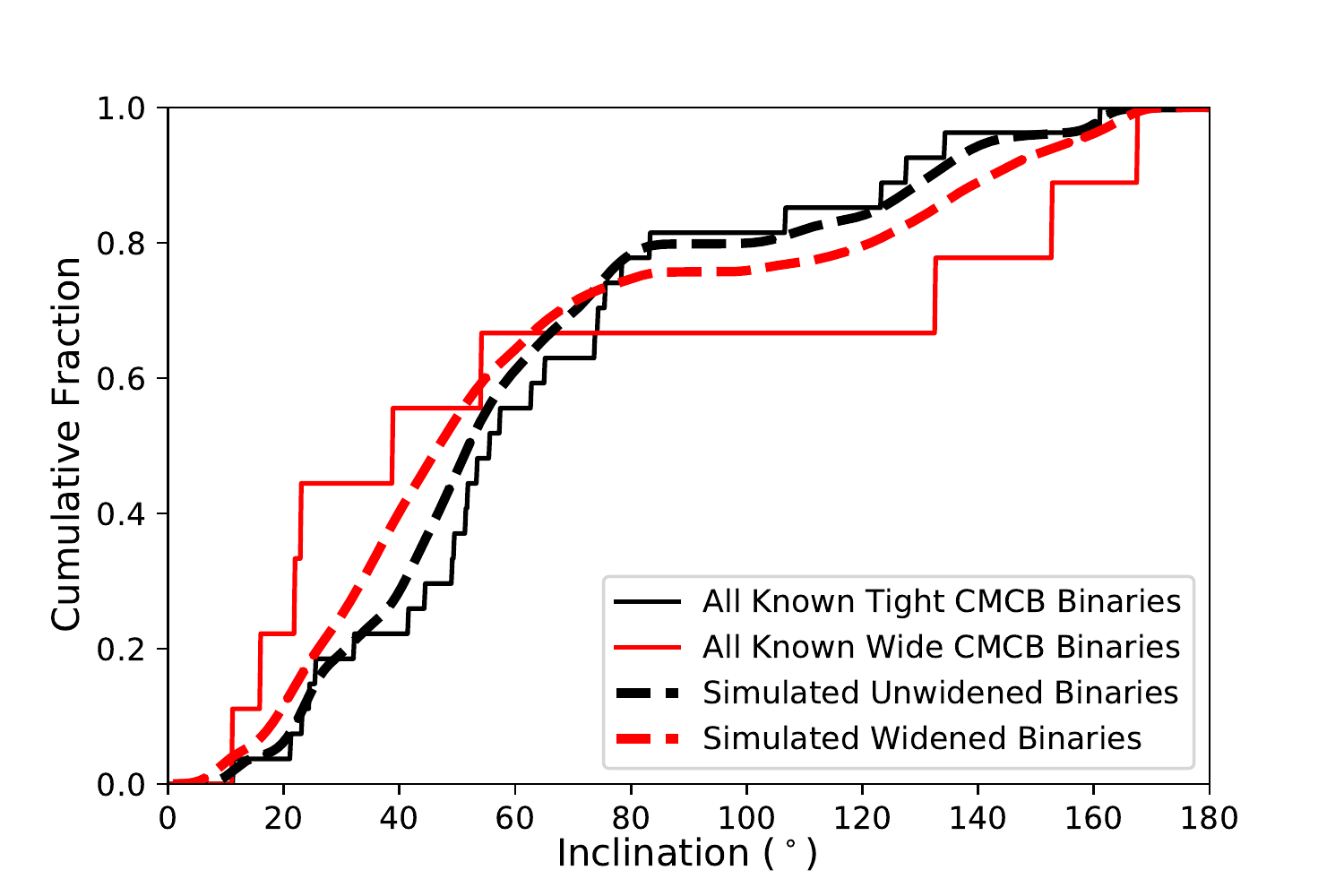}
\caption{The inclinations of our simulated binaries that have become Ultra-Wide vs those that have not.  These compared to the inclinations of observed Ultra-Wide binaries and tight binaries.}
\end{figure}

With all of our SFDs, a K-S statistic cannot rule out the possibility that the known Ultra-Wide TNB inclination distribution is drawn from our gravitationally evolved Ultra-Wide TNBs.  However, with the low number of Ultra-Wide TNBs known at present, a K-S test similarly cannot rule out the possibility of the known wide inclination sample being underivable from the known tight sample.  This means that while our widened inclination distribution does not conflict with the observed sample, its similarity is scarcely more significant than the tight binary sample.  

While we have been unable to replicate the number of Ultra-Wide TNBs that we presently see, it seems that binaries widened through gravitational perturbations have orbital properties similar to those observed.  Our widened eccentricity distribution greatly resembles the observed sample while our inclination distribution is not in conflict with it.  

\section{Summary}

We have presented an in-depth analysis into the effects of the gravitational interactions of passing bodies on Kuiper Belt binaries.  

\begin{enumerate}
  \item Gravitational interactions are a significant contribution, comparable to collisional interactions, in the evolution of TNBs.  If subjected to the gravitational encounters of the modern Kuiper Belt, the binaries 2001 QW$_{322}$ and 2000 CF$_{105}$ have average lifetimes comparable to the age of the solar system.
  
  \item Through gravitational evolution, Ultra-Wide TNBs can be gradually formed from tighter binaries.  However, with even our most optimistic of initial conditions, we cannot form wide binaries in the abundances we would expect.  Its possible though, that accounting for the historical declining population of the Kuiper Belt itself would allow larger populations of wide binaries to form.
  
  \item If such widened binaries begin with inclinations and eccentricities sampled from known tight TNBs, their eventual widened orbital properties are consistent with those of observed Ultra-Wide TNBs.  Our widened eccentricity distribution in particular greatly resembles the known distribution of Ultra-Wide TNBs.  Thus, binaries widened through gravitational perturbations are not dissimilar to those that we observe.
   
\end{enumerate}

\section{Acknowledgements}
This work was performed with support from NASA Emerging Worlds grant 80NSSC18K0600.  Our computing was performed at the OU Supercomputing Center for Education \& Research (OSCER) at the University of Oklahoma (OU).

\section{Data Availability}

The data underlying this article will be shared on reasonable request to the corresponding author.

\bibliographystyle{aasjournal}
\bibliography{cites} 

\end{document}